\documentclass[prd,preprint,eqsecnum,nofootinbib,amsmath,amssymb,
               tightenlines,dvips]{revtex4}

\usepackage[dvips]{graphicx}
\usepackage{bm}

\def\x{{\bf x}}
\def\y{{\bf y}}
\def\z{{\bf z}}
\def\p{{\bf p}}
\def\k{{\bf k}}
\def\q{{\bf q}}
\def\v{{\bf v}}

\def\E{{\bf E}}

\def\Im{{\rm Im}}

\def\x{{\bf x}}

\def\Nc{N_{\rm c}}

\def\st{\begin{equation}}
\def\stp{\end{equation}}
\def\bg{\begin{eqnarray}}
\def\nd{\end{eqnarray}}
\def\Eq#1{Eq.~(\ref{#1})}
\def\app#1{Appendix~\ref{#1}}
\def\Fig#1{Fig.~\ref{#1}}
\def\Sect#1{Section~\ref{#1}}

\def\llangle{\left\langle}
\def\rrangle{\right\rangle}

\def\EM{{\scriptscriptstyle EM}}

\def\tr{\operatorname{tr}}

\def\dpslash{\frac{d^3\p}{(2\pi)^3} \>}

\def\Mj{M_{\scriptscriptstyle J/\psi}}
\def\JJ{{\scriptscriptstyle JJ}}
\def\NN{{\scriptscriptstyle NN}}
\def\ppx#1{\frac{\partial}{\partial #1} }
\def\Fv{f_{\scriptscriptstyle V}}

\def\gsim{\mbox{~{\protect\raisebox{0.4ex}{$>$}}\hspace{-1.1em}
	{\protect\raisebox{-0.6ex}{$\sim$}}~}}
\def\lsim{\mbox{~{\protect\raisebox{0.4ex}{$<$}}\hspace{-1.1em}
	{\protect\raisebox{-0.6ex}{$\sim$}}~}}
\def\jpsi{{J/\psi}}

\def\nott#1{\setbox0=\hbox{$#1$}                
   \dimen0=\wd0                                
   \setbox1=\hbox{/} \dimen1=\wd1               
   \ifdim\dimen0>\dimen1                        
      \rlap{\hbox to \dimen0{\hfil/\hfil}}      
      #1                                        
   \else                                        
      \rlap{\hbox to \dimen1{\hfil$#1$\hfil}}   
      /                                         
   \fi}                                         %

\def\dndep{\frac{\partial n}{\partial E_{\bf p}} }

\advance\parskip 1.9pt
\advance\voffset -0.2in

\begin{document}

\title{Heavy Quark Diffusion from the Lattice}

\author{P\'eter Petreczky}
\affiliation
    {%
    Nuclear Theory Group, 
    Department of Physics,
Brookhaven National Laboratory, 
Upton, New York 11973, USA
    }%
\author{Derek Teaney}
\affiliation
    {%
    Department of Physics \& Astronomy,
    SUNY at Stony Brook,
    Stony Brook, New York 11764, USA
    }%

\date{\today}

\begin{abstract}
We study the diffusion of heavy quarks in the Quark Gluon 
Plasma using the Langevin equations of motion and estimate
the contribution of the transport peak to the Euclidean
current-current correlator. We show that the Euclidean
correlator is remarkably insensitive to the heavy quark 
diffusion coefficient and give a simple physical interpretation
of this result using the free streaming Boltzmann equation. However if 
the diffusion coefficient is smaller than $\sim 1/(\pi T)$, as favored
by RHIC phenomenology, the transport contribution should be visible
in the Euclidean correlator. We outline a procedure 
to isolate this contribution.

\end{abstract}



\maketitle


\section {Introduction}

The experimental relativistic heavy ion program has produced a variety
of evidences which suggest that a Quark Gluon Plasma (QGP) has been
formed at the Relativistic Heavy Ion Collider (RHIC)
\cite{Bellwied:2005kq,Adcox:2004mh}.  One of the most exciting results
from  RHIC  so far is the large azimuthal anisotropy of light hadrons
with respect to the reaction plane, known as elliptic flow.  The
observed elliptic flow  is significantly larger than was expected from
kinetic calculations \cite{Molnar:2001ux}, but in fairly good agreement
with simulations based upon ideal hydrodynamics
\cite{Ollitrault:1992bk,Hirano:2004er,Teaney:2001av,Kolb:2000fh,Huovinen:2001cy}.
This result suggests that  the transport mean free path is small
enough to employ thermodynamics and hydrodynamics to describe the heavy
ion reaction. However, this interpretation of the RHIC results demands
further theoretical and experimental corroboration.

Experimentally, this interpretation can be challenged by measuring the
elliptic flow of  charm and bottom mesons \cite{Laue:2004tf,Adler:2005ab,Adler:2004ta}.  
The first experimental results show a non-zero elliptic 
flow for these heavy mesons.
Naively, since the
quark mass is significantly larger than the temperature of the medium,
the relaxation time of heavy mesons is $\sim M/T$ longer
than the light hadron relaxation time
\[
      \tau_R^{\rm heavy} \sim \frac{M}{T} \tau_R^{\rm light} \, .
\]
Consequently the heavy meson elliptic flow should be reduced relative
to the light hadrons.  Recently, a variety of phenomenological
models have estimated how the transport mean free path of heavy
quarks in the medium is ultimately reflected in the
elliptic flow \cite{Moore:2004tg,Molnar:2004ph,Zhang:2005ni}. 
The result of these model studies is best expressed in terms of the 
heavy quark diffusion coefficient. 
(In a relaxation time approximation the diffusion coefficient is related
to the equilibration time, $ \tau_R^{\rm heavy}=\frac{T}{M}D$.)
There is a consensus from the models that if the diffusion coefficient of the heavy
quark is greater than 
\[
D \gsim \frac{1}{T} \,,
\]  
the heavy quark elliptic
flow will be small and probably in contradiction with current data. 

Theoretically,  transport coefficients have
been computed in the perturbative quark gluon plasma using 
kinetic theory \cite{Baymetal,AMY6}. The heavy quark diffusion
coefficient has also been computed \cite{Moore:2004tg, Svetitsky:1987gq,Braaten:1991we}. Recent efforts have also explored
some meson resonance models and found a substantially smaller 
diffusion coefficient than in perturbation theory \cite{vanHees:2004gq}.
The ambiguity in these calculations underscores the need 
for reliable  non-perturbative estimates of transport  coefficients
in the QGP.

Kubo formulas relate hydrodynamic 
transport coefficients  to the small frequency behavior
of real time correlation functions \cite{Forster,BooneYip}.
Correlation functions in real
time are in turn related to correlation functions in imaginary time by analytic
continuation. Karsch and Wyld \cite{karschwyld} 
first attempted to use this connection
to extract the shear viscosity of QCD from the lattice.
More recently, additional attempts to extract the shear
viscosity \cite{nakamura97,nakamura05} and electric conductivity \cite{gupta03} have been made.
We will argue on general grounds that Euclidean correlations functions 
are remarkably insensitive 
to transport coefficients. For weakly coupled field theories 
this has been discussed by Aarts and Resco \cite{aarts02}.
For this reason, only precise lattice data and a comprehensive
understanding of the different contributions to the Euclidean
correlator can constrain the transport coefficients.

In this paper we are going to estimate the contribution of heavy quark
diffusion to Euclidean vector current correlators. 
The case of heavy quarks is special since  the time scale for
diffusion, $M/T^2$, is much longer than any other time scale in the problem.
In terms of the spectral functions, this separation 
means that transport processes
contribute at small energy, $\omega \sim T^2/M$,
and all other contributions (e.g. resonances and continuum contributions) 
start 
at high energy, $\omega \gsim 2 M$. For light quarks, transport contributes to meson spectral functions for $\omega \sim g^4 T$. This 
scale is 
separated from the energy scale of other contributions, $\omega \sim T, g T$, only in the weak coupling limit $g \ll 1$.

The behavior of vector current correlators at large times can be
related to the heavy quark diffusion constant.
Euclidean heavy meson correlators at temperatures above the
deconfinement temperature have been  calculated on the lattice and
attempts to extract spectral functions have been made
\cite{umeda02,asakawa04,datta04}. Transport should show up as a peak at
very small frequencies, $\omega \simeq 0$. So far, it has not been
observed in these studies. Obviously, it is very difficult to
reconstruct the spectral functions from the finite temperature lattice
correlators, as the time extent is limited by the inverse temperature.
However, the temperature dependence of the correlators can
be determined to very high accuracy \cite{datta04,jako05} and therefore
some information about the transport can be ascertained.

\section{Linear Response and the Spectral Density}
\label{linear_response_section}

This  section briefly reviews linear response  
which is the 
appropriate framework to connect the Langevin and diffusion
equations to the current-current correlator \cite{Forster}.  We will
also define the spectral density which is needed 
to relate the Euclidean current-current correlator
measured on the lattice to its Minkowski counterpart.

Consider a small  
perturbing Hamiltonian
\st
    H = H_0 - \int d^3\x \, h(\x, t)\, O(\x, t) \, ,
\stp
where $h(\x,t)$ is a classical source.
Now imagine that we slowly turn on the external source 
$h(\x,t)$, and then abruptly turn it off at time $t=0$. $h(\x,t)$ obeys
\st
\label{fields}
    h(\x,t) = e^{\epsilon t}\theta(-t) \, h^{0}(\x)   \,.
\stp
The expectation value of $\llangle \delta O(\x,t) \rrangle$ in 
the presence of the perturbing Hamiltonian is
\begin{equation}
   \llangle \delta O(\x, t) \rrangle =  +i \int d^3\y \int_{-\infty}^{t}dt'\, 
   \llangle \left[ O(\x, t), O(\y, t') \right] \rrangle\ h(\y, t') \, .
\end{equation}
Using translational invariance and taking spatial Fourier transforms
we have
\begin{equation}
   \llangle \delta O(\k, t) \rrangle = \int_{-\infty}^{+\infty}dt'\,
\chi(\k, t-t') \, h(\k,t') \,,
\end{equation}
where 
\begin{equation}
\label{pttheory}
   \chi(\k, t-t') = 
\int d^3\x \> e^{-i\k\cdot\x}\, i\theta(t-t')\,\llangle \left[ O(\x, t), O(\y, t') \right] \rrangle \, ,
\end{equation}
is the retarded correlator.
When confusion can not arise we use momentum labels 
$\p,\k,\q,\dots$ rather than position labels
$\x,\y,\z,\dots$ to distinguish the spatial Fourier transform of a 
field $\llangle O(\k,t)\rrangle=\int e^{i\k\cdot \x} \llangle O(\x,t)\rrangle$ from the field itself, $\llangle O(\x,t) \rrangle$.

For $t > 0$, differentiating with respect to $t$ we have 
\st
  \ppx{t} \llangle \delta O(\k,t) \rrangle
    = \int_{-\infty}^{+\infty} dt'\, \ppx{t} \chi(\k, t-t') 
    \,  h(\k,t') \,.
\stp
Using $\ppx{t} \chi(\k,t-t') = -\ppx{t'}\chi(\k,t-t')$,
 integrating by parts with respect to $t'$, and using \Eq{fields}, 
we find a relation between expectation values and correlators
\st
\label{linear_response}
 \ppx{t} \llangle \delta O(\k,t) \rrangle = - 
  \chi(\k,t) h^0(\k) \, .
\stp
The external field $h^0(\k)$ can be eliminated by using  
the relation between the static susceptibility $\chi_s$, 
the initial condition $\llangle \delta O(\k,t) \rrangle$,
and the external field 
\st
\label{static_succeptibility}
   \llangle \delta O(\k,t=0) \rrangle = \chi_s(\k) \, h^{0}(\k) \, ,
\stp
where the static susceptibility $\chi_s(\k)$, follows from \Eq{pttheory}
\st
\label{chis}
   \chi_s(\k) = \int_{0}^{\infty} dt'\, e^{-\epsilon t'}\, \chi(\k,t') \, .
\stp
Eliminating the field $h^0(\k)$, we find 
\st
\label{linear_response_nofield}
 \chi_s(\k) \ppx{t} \llangle \delta O(\k,t) \rrangle = - 
  \chi(\k,t) \, \llangle \delta O(\k,t=0) \rrangle  \, .
\stp
This result relates the time evolution of an average 
from a specified initial condition to 
an equilibrium correlator $\chi(\k,t)$.

The function $\chi(\k,t)$ is related to the spectral 
density. The Fourier transform of the retarded correlator
can be written 
\st
\label{chi}
   \chi(\k, \omega) = \int_{0}^{+\infty} dt\, e^{+i\omega t} \,
     \chi(\k,t) \,.
\stp
$\chi(\x,t)$ is real, and since the integration is only over positive times, $\chi(\k,\omega)$ is analytic in the upper half plane.
Provided the Hamiltonian is time-reversal
invariant and the operator $O$ has definite signature under time reversal, $\llangle [O(\x,t)\,,O(\y,0)]\rrangle$ is an
odd function of time and $\chi(\k,t)$ is an even (odd) function of
$\k$ (time).
The
spectral density,  $\rho(\k,\omega)$, is defined as the
imaginary part by $\pi$ of the retarded correlator 
\st
\label{spectral_density}
  \rho(\k,\omega) = \frac{\Im \chi(\k,\omega)}{\pi}  = 
         \frac{1}{2\pi} \int d^3\x \int_{-\infty}^{\infty}
             e^{-i\k\cdot\x + i\omega t} 
           \llangle [ O(\x,t), O(0,0) ] \rrangle \, .
\stp
By inserting complete sets of states, one may show that 
the spectral density is an odd function of 
frequency and is positive for $\omega >0$ \cite{leBellac}.

The Euclidean correlator may be deduced from the spectral 
density. 
Euclidean tensors  are defined from their Minkowski
counter parts, 
$\left. O_{M}
^{\mu_1\dots\mu_n}\right._{\nu_1\dots\nu_n} (-i\tau)\equiv(-i)^r (i)^s \left. O_{E}^{\mu_1\dots\mu_n}\right._{\nu_1\dots\nu_n}(\tau)$,
where $r$ and $s$ are the number of zeros in $\{\mu_1\dots\mu_n\}$ 
and $\{\nu_1\dots\nu_n\}$ respectively.
In what follows, we will drop the ``$M$''  on Minkowski operators but indicate
``$E$'' on Euclidean operators.
With these definitions $x^{0}=-ix^{0}_{E}=-i\tau$, 
and Euclidean tensors transform under $O(4)$ in the 
zero temperature limit. 
Correlators in Euclidean space time are of the following form:
\st
   G(\k,\tau) = \int d^3\x\, e^{i\k \cdot \x} \,\llangle O_{E}(\x, \tau)\,
 O_{E}(0, 0) \rrangle  \equiv (-1)^{r+s} \int d^3\x\, e^{i\k \cdot \x}\, D^{>}(\x,-i\tau)\,,
\stp
where  $D^>(\x,t) = \llangle O(\x,t) O(0, 0) \rrangle $\,.
Usually,  the lattice works with at zero spatial momentum $\k=0$.
In Minkowski space, we work 
with the Fourier transform  of $D^{>}(\x,t)$,
\st
\label{Dgreater}
D^{>}(\k,\omega) =  
\int d^4x\, e^{+i\omega t - i\k\cdot\x} \,D^{>}(\x,t)  \, .
\stp
Similarly, we define $D^{<}(\x,t) \equiv \llangle O(0,0)\, 
O(\x,t) \rrangle$ and its Fourier transform.
Thus, the spectral density, \Eq{spectral_density},
is given by
\st
   \rho(\k,\omega)  = \frac{D^{>}(\k, \omega) - D^{<}(\k, \omega) }{2\pi} \, .
\stp 
Using the Kubo-Martin Schwinger (KMS) relation 
$D^{>}(\k,t) = D^{<}(\k,t + i/T)$, and its Fourier counter-part
$D^{>}(\k,\omega) = e^{+\omega/T} D^{<}(\k,\omega)$,
one discovers the relation between the spectral density 
and the Euclidean correlator,
\st
  G(\k,\tau) = (-1)^{r+s} \int_0^{\infty} d\omega \, \rho(\k,\omega) \frac{ \cosh\left( \omega\left(\tau -\frac{1}{2T}\right)\right) }
{\sinh\left(\frac{\omega}{2T}\right) } \, .
\label{spectral_rep}
\stp
Again, given an operator,   $\left. O ^{\mu_1\dots\mu_n}\right._{\nu_1\dots\nu_n}$, \,
$r$ and $s$ are the number of zeros in the space-time indices $\{\mu_1\dots\mu_n\}$ and $\{\nu_1\dots\nu_n\}$ respectively.

For our discussion two correlators will be important:
the density-density correlator 
\begin{equation}
D^{>}_{\scriptscriptstyle NN}({\bf x},t)=\llangle J^0({\bf x},t) J^0(0,0) \rrangle\,,
\end{equation}
and the current-current correlator
\begin{equation}
D^{>,ij}_{\scriptscriptstyle JJ}({\bf x},t)=\llangle J^i({\bf x},t) J^j(0,0) \rrangle\,.
\end{equation}
These correspond to the Euclidean correlators calculated on the
lattice
\begin{equation}
G_\NN({\bf x},\tau)=\llangle J^0_{E}({\bf x},\tau) J^0_{E}(0,0) \rrangle=
-D_\NN^>({\bf x}, -i \tau) \, ,
\end{equation}
\begin{equation}
G_\JJ^{ij}({\bf x},\tau)=\llangle J^i_{E}({\bf x},\tau) J^{j}_E(0,0) \rrangle=
D_\JJ^{>,ij}(\x, -i \tau) \, .
\end{equation}
The corresponding retarded correlators $\chi_\NN({\bf x},t)$
and $\chi^{ij}_\JJ({\bf x},t)$ can be introduced in the same way.
The Fourier transforms of current-current correlators can be 
decomposed into longitudinal and transverse parts. For
the retarded correlator we write:
\st
\label{chiL}
   \chi_\JJ^{ij}(\k,\omega) = \left(\frac{k^i k^j}{k^2} - \delta^{ij}\right)\, \chi_\JJ^T(\k,\omega)
   + \frac{k^i\,k^j}{k^2}\, \chi_\JJ^L(\k,\omega) \, .
\stp
Current conservation relates the density-density and the longitudinal
current-current correlators
\st
\label{chiJJ}
\frac{\omega^2}{k^2} \chi_{\scriptscriptstyle NN}(\k,\omega) = \frac{k^ik^j}{k^2} \chi_\JJ^{ij}(\k,\omega) = \chi_\JJ^L(\k,\omega)\,.
\stp
For $k=0$ there is no distinction between the longitudinal and transverse parts
and therefore  for $k \ll T$ ,  $\chi_\JJ^L(\k,\omega) \simeq
\chi_\JJ^T(\k,\omega)$. Since the transverse component of 
the current-current correlator is not studied in this 
work, we will drop the ``L",   
and for instance, $G_\JJ$ and $\rho_\JJ$ are 
short for $G_\JJ^{L}$ and $\rho_\JJ^L$.

At finite temperature the spectral function can be written as 
\st
   \rho_\JJ(\k, \omega) = \rho_\JJ^{\rm low}(\k,\omega) + \rho_\JJ^{\rm high} (\k,\omega)\,,
\stp
where the last term is just the zero temperature part and the first term is the low energy $\omega \ll T^2/M$ 
contribution.  In the next two sections we will discuss how to estimate the 
low frequency part.

\section{Transport in Euclidean Correlators}
\label{transport_in_euclid}

In this section we estimate how the low frequency part
of the spectral function contributes to 
the Euclidean current-current correlator.
To leading order, the 
moments of the spectral function, the time
derivatives of the retarded correlator  at $t=0$, and
the derivatives of the Euclidean correlator at $\beta/2$,  
are in one-to-one correspondence.
Since the leading contribution is
due to time derivatives at $t=0$,  the free streaming of
heavy quarks gives the dominant contribution to the Euclidean
current-current correlator. The first scattering
correction appears in the second (fourth) derivative at $\beta/2$ 
of the current-current (density-density) euclidean correlator.

First let us start with the density-density correlator. For small frequencies, the kernel
in Eq. (\ref{spectral_rep}) is given by $2 T/\omega$, and thus we can write
\st
-G_\NN^{\rm low}({\bf k},\tau) \simeq 2T\,\int_0^{\infty} \frac{d\omega}{\omega}\, \rho_\NN^{\rm low}(\k,\omega) \, .
\stp
Inserting the definition of the spectral density
\st
\rho_\NN^{\rm low}(\k,\omega)=\frac{1}{\pi} \int_0^{\infty} dt \,\sin(\omega t) \,\chi_\NN(\k,t) \, , 
\stp
and performing the integral over frequency, we find
\st
-G_\NN^{\rm low}(\k,\tau) \simeq \int_0^{\infty} dt\, \chi_\NN^{\rm low}(\k,t)=T \chi_s(\k) \,.
\stp
The last equality follows from \Eq{chis}.
 Similarly, the low energy contribution to the longitudinal current correlator is 
\st
G_\JJ^{\rm low}({\bf k},\tau) \simeq 2T \int_0^{\infty} \frac{d \omega}{\omega} 
\,\rho_\JJ^{\rm low}(\k,\omega) \,
\left[1 - \frac{1}{6}\left(\frac{\omega}{2T}\right)^2 + \omega^2\,\frac{1}{2} \left(\tau-\beta/2\right)^2 + \dots\right] \, .
\label{GJJmoment}
\stp
Inserting the spectral density
\st
\rho_\JJ^{\rm low}(\k,\omega)=\frac{\omega^2}{\pi\, k^2}  \int_0^{\infty} dt\, \sin(\omega t)\, \chi_\NN(\k,t) \, ,
\stp
and performing the integral over frequency, we find
\st
G_\JJ^{\rm low}(\k,\tau)=\frac{T}{k^2} \left[ \partial_t^{(1)} \chi_\NN(\k,t)+
\frac{1}{24\,T^2}\partial_t^{(3)}\chi_\NN(\k,t) 
\, -\,
\partial_t^{(3)}\chi_\NN(\k,t) 
\, \frac{1}{2} \left(\tau-\beta/2\right)^2 
+ \dots \right]_{t=0} \, .
\label{GJJ}
\stp
Thus we see that the dominant low energy contributions to the Euclidean
correlator is given by the short time behavior of the retarded
correlator.  Indeed, as seen from \Eq{GJJmoment} and \Eq{GJJ}, the moments of
the spectral function, the $\tau$ derivatives of the Euclidean
correlator at $\beta/2$, and the time derivatives of the real-time
retarded correlator at $t=0$ are in one to one correspondence. 
While a short time expansion can never be used to rigorously extract transport
coefficients,  they have proved useful in 
in non-relativistic contexts \cite{Forster,BooneYip}

For times which are short compared to the collision time
it is reasonable to expect that the motion of heavy quarks
is described by the free-streaming Boltzmann equation. 
Even in the interacting theory, the free streaming Boltzmann equation will describe the 
first time derivative of the retarded correlator or 
the first term in the Euclidean correlator, \Eq{GJJ}.

Let us  create an 
excess of heavy quarks,  
and subsequently study the diffusion 
of this excess at short times.
This can be done by introducing 
a small chemical 
potential $\mu(\x) = \mu_0 + \delta \mu(\x)$ 
as in \Sect{linear_response_section}. 
Then
the thermal distribution function  at an initial time $t=0$ is
\st
   f_0(\x, \p) \equiv \frac{1}{e^{ (E_p -\mu(\x))/T } \mp 1} \approx f_p + f_p\,(1 \pm f_p) \,\frac{ \delta\mu(\x) }{T} \, ,
\stp
with\footnote{
Generally we will restrict ourselves to a 
heavy quark limit where there are well defined
high and low frequency contributions. The discussion in this
paragraph and the previous paragraph applies whenever the
scale separation persists, and is therefore applicable to 
relativistic weakly coupled quarks . We will therefore 
generalize this paragraph to relativistic quarks with Bose-Einstein and Fermi-Dirac statistics. },\, $f_p = 1/(e^{(E_p -\mu_0)/T} \mp 1)$\,. For short 
times the collision-less Boltzmann equation applies, 
\st
   \left[ \ppx{t} + v_\p^{i} \ppx{x^{i}} \right] f(\x,\p,t) = 0 \,.
\stp
The solution to this equation with the specified initial conditions 
is
\st
\label{collisionless}
   f(\x,\p,t) = f_0(\x - v_\p t, \p)  \,.
\stp
Then
the fluctuation in the number density is
\st
  \delta N(\x,t)  = \int \dpslash \, \delta f(\x,\p,t) \, ,
\stp
with  $\delta f(\x,\p,t) = f(\x,\p,t) - f_p$\;. 
Then taking spatial Fourier transforms with $\k$ conjugate to $\x$ 
and substituting the distribution function, \Eq{collisionless}, we have
\st
\label{freeN}
  \delta N(\k,t)  = \left[\frac{1}{T} \int \dpslash e^{-i\k \cdot v_\p t} f_p(1\pm f_p)\right] \delta \mu(\k) \, .
\stp  
For small times, we expand the exponential, and find
\st
\label{NNt}
  \delta N(\k,t)  =  \left[\chi_s(\k) - \frac{1}{2}\,t^2\,k^2\,\chi_s(\k) \llangle \frac{v^2}{3}\rrangle  \right] \delta \mu(\k) \, ,
\stp  
with 
\st
   \chi_s(\k) = \frac{\partial N}{\partial \mu_0} =
\frac{1}{T} \int \dpslash f_p(1\pm f_p)  \, ,
\label{chis_boltzmann}
\stp
and
\st
  \llangle \frac{v^2}{3} \rrangle = 
 \frac{1}{T\chi_s(\k)} \,\int \dpslash f_p(1\pm f_p) \,\frac{v_\p^2}{3}  \, .
\label{v2by3}
\stp
Thus, from \Eq{GJJ}, \Eq{linear_response}, and \Eq{NNt},  
we find\footnote{
Here we have considered only a single component gas. For 
the case of heavy quark diffusion, \Eq{chis_boltzmann} and \Eq{v2by3} should
be multiplied by $4\Nc$ to account for the sum over spin, color, and 
and anti-quarks.}
\st
\label{GJJ_kinetic_theory}
   G_\JJ^{L, \,\rm low}(\k,\tau) = T\chi_s(\k)\, \llangle \frac{v^2}{3} \rrangle \, .
\stp
In the free theory, at $\k=0$ there there are no corrections
to this result  and the Euclidean correlator is a constant. At
finite $\k$, the lattice  correlator is not a constant even in 
the free theory.
For massless particles, $\llangle v^2/3 \rrangle = 1/3$, while for massive we have
$\llangle v^2/3 \rrangle = T/M$.  

We have outlined the short time expansion of $\chi_\NN(\k,t)$ .
Further insight is gained from the full free spectral function.
From, \Eq{freeN}, \Eq{linear_response}  and a simple Fourier transform we deduce that the retarded correlator
from the free streaming Boltzmann equation is 
\[
   \chi_\NN(\k,\omega) = \frac{1}{T}\int \dpslash f_p(1\pm f_p) \,
       \frac{-\k\cdot\v_\p}{\omega - \k\cdot\v_\p + i\epsilon } \, .
\] 
Taking the imaginary part, 
the corresponding spectral density is 
\[
   \rho_\NN^{\rm low}(\k,\omega) = \frac{1}{T}\int \dpslash f_p(1\pm f_p) 
\, \k \cdot \v_\p\, \delta(\omega - \k\cdot\v_\p) \, .
\]
As shown in \app{loopspec},
this form for the spectral density is identical to the 
one loop spectral function of the free theory at small 
$k$ and $\omega$, \Eq{loop_transport}. As discussed in \app{loopspec},
the resulting integral can be performed in the non-relativistic
limit and we find
the free spectral function for the heavy quark current-current 
correlator
\st
   \rho_\JJ^{\rm low}(\k, \omega)  
=  \chi_s \, 
    \frac{\omega^3}{k^2}\frac{
1} {\sqrt{2\pi k^2 \llangle \frac{v^2}{3} \rrangle } }
\exp\left(-
\frac{\omega^2}{2k^2\llangle \frac{v^2}{3}\rrangle}\right)   \, .
\label{free_rhojj_k}
\stp
This is the dynamic structure factor of a free non-relativistic gas \cite{BooneYip}.
In the free theory, the spectral function 
is essentially a Gaussian, with a width that is 
proportional to $k^2$.  In the limit that $\k=0$ the
correlator is 
\st
   \rho_\JJ^{L,\rm low}(\k, \omega) =  \chi_s \llangle \frac{v^2}{3} \rrangle \,\omega \delta(\omega) \, .
\label{free_rhojj_k0}
\stp

In the free theory, the low frequency spectral
density is infinitely narrow at $\k=0$.  The moments of the spectral
density are in one to one correspondence with the derivatives of the
Euclidean correlator at  $\beta/2$. Since higher moments of a delta
function are zero,  all derivatives at $\beta/2$ vanish and the low
frequency contribution of the free theory to the Euclidean correlator
is simply a flat line. Thus, provided the high frequency contribution
of the spectral function can be subtracted,  any bending of the
Euclidean correlator is indicative of something beyond free streaming.
In the next sections we will discuss how interactions smear the 
$\delta(\omega)$ function and estimate 
how much the Euclidean correlator curves at $\beta/2$ as
a function of diffusion coefficient. 

\section{Heavy Quark Diffusion in the Langevin Effective Theory}
\label{langevin_section}

In this section we will discuss the predictions of the 
Langevin equations for the  retarded correlator. 
As mentioned before, the time scale for heavy quark transport, $M/T^2$
is much larger than typical time scale for light degrees of freedom
in the plasma. For this reason
we will assume that the Langevin equations provide a good 
macroscopic description of the thermalization of charm quarks \cite{Moore:2004tg},
\begin{eqnarray*}
\label{newton_langevin}
\frac{dx^i}{dt} &=& \frac{p^i}{M}\,,  \nonumber \\
\frac{dp^i}{dt} &=& \xi^i(t) - \eta p^i\,, \qquad \langle \xi^i(t) \xi^j(t') \rangle = \kappa \delta^{ij}
	\delta(t-t') \, .
\end{eqnarray*}
The drag and fluctuation coefficients are related by the 
fluctuation dissipation relation 
\begin{equation}
\label{etad}
\eta = \frac{\kappa}{2 MT} \, .
\end{equation}

For timescales which are much larger than $1/\eta$ the 
heavy quark number density obeys ordinary diffusion equation
\[
   \partial_t N + D \nabla^2 N = 0 \,.
\]
The drag coefficient $\eta$ can be related to the diffusion coefficient through the Einstein relation
\begin{equation}
   D = \frac{T}{M \eta} = \frac{2 T^2}{\kappa}  \, .
\label{eq:D}
\end{equation}

The effective Langevin theory can be derived from 
kinetic theory in the weak coupling limit \cite{Moore:2004tg}
and probably is
adequate for describing heavy quark diffusion even for
strongly interacting plasma.
The Langevin equations make a definite prediction for
the retarded correlator. 
Following the framework of linear response,
consider an initial distribution
of heavy quarks when a small perturbing chemical potential is applied,
$\mu(\x) = \mu_0 + \delta \mu(\x)$.
The initial phase space distribution of heavy quarks is
\begin{equation}
   f(\x,\p,t=0) = e^{\frac{\mu(\x)}{T}-\frac{M}{T}}
e^{ - \frac{p^2}{2 M T} } \, .
\end{equation}
Summing over spins and colors, the initial number density 
of quarks minus anti-quarks is 
\st
\label{number_density}
   N(\x,t=0) = \left[ 4 N_c\right]\, \left(\frac{M\,T}{2\pi}\right)^{3/2} \, e^{-\frac{M}{T}} \,
 \sinh\left(\frac{\mu(\x)}{T}\right) \,.
\stp
By comparing \Eq{number_density} and \Eq{static_succeptibility},
we find the static susceptibility
\st
\label{explicit_static}
    \chi_s =  [4N_c] \,\left(\frac{M\,T}{2\pi}\right)^{3/2}\,  e^{- \frac{M}{T}}  \, \cosh\left(\frac{\mu_0}{T}\right) \, .
\stp

Let $P(\x, t)$ be the probability that
a heavy quark starts at the origin at $t=0$  and moves a distance 
$\x$ over a time $t$ .
Consider the relaxation of an initial distribution 
of heavy quarks $N(\x,t=0)$ slightly perturbed from equilibrium. 
The distribution of heavy quarks at a later time is,
\st
    N(\x,t) = \int\,d^3\x'\, P(\x - \x', t)\, N(\x',0) \,,
\stp
or
\st
   N(\k, t) = P(\k, t) N(\k, 0)  \, .
\stp 
Comparing this result with the linear response result, \Eq{linear_response_nofield}, 
we conclude that for small $\k$ and times large compared to 
typical medium timescale 
\st
\label{Pchirelation}
  \chi_\NN(\k, t)
     = -\chi_s(\k) \,\partial_t  P(\k, t)     \,.
\stp
Thus, to find the retarded correlator $\chi_\NN(\k,\omega)$ ,
we need only find the probability $P(\x, t)$. 

The probability distribution $P(\x, t)$ is determined in \app{brownian}.
Not surprisingly, the distribution 
is a Gaussian,
\st
\label{gaussian}
   P(\x, t) = \frac{1}{(2\pi \sigma^2(t))^{3/2}} 
              \exp\left( -\frac{1}{2}\frac{x^2}{\sigma^2(t)} \right)  \,,
\stp
with a  width that depends non-trivially on time
\st
\label{gaussian_width}
    \sigma^2(t) = 2 D t - \frac{2 D}{\eta}(1 - e^{-\eta t}) \,.
\stp
For large times, we have  $\sigma^2(t) \approx 2 D \,t$ as expected
from the ordinary diffusion equation. For small times, we 
have 
\st
\sigma^2(t)\approx (T/M)\, t^2    \qquad\qquad (\eta t \ll 1) \,,
\stp
which reflects the initial thermal velocity distribution of heavy quarks,
$\llangle v^2/3 \rrangle = T/M$.
Using \Eq{Pchirelation}, the probability distribution 
\Eq{gaussian}, and the definition of
the retarded correlator, we find the following form:
\begin{equation}
\label{chiNN}
    \chi_\NN(\k,\omega) = \chi_s(\k) \, \int_{0}^{\infty}dt\, e^{i\omega t}
 \, k^2D\, (1 - e^{-\eta\, t})\, e^{-k^2 D t\  +\  (k^2 D/\eta)\, (1-e^{-\eta\, t})}  \, .
\end{equation}
\Eq{chiNN} summarizes the contribution of
the Langevin equations to the retarded density-density correlator.
The retarded correlator has following properties:  
\begin{itemize}

\item[1.] 
For small $\k$,  $Dk^2 \ll \eta$, and arbitrarily large times, we may write the integrand
as $ k^2D\, (e^{-k^2 D t} - e^{-\eta\, t})$, and perform
the integration
\st
    \chi_\NN(\k,\omega) =  \frac{\chi_s(\k) \, D k^2 }{-i\omega + k^2 D}
                  - \frac{\chi_s(\k) \,D k^2}{-i\omega + \eta}
\,.
\stp
For small frequency $\omega \sim Dk^2$, the first term dominates and
recalls the diffusion equation, $(\partial_t + D\nabla^2)^{-1}$. For 
large frequencies $\omega \sim \eta$,  $\chi_{NN}$  recalls the 
drag term of the Langevin equations,  $(\partial_t + \eta)^{-1}$ .
Of particular relevance to lattice measurements is 
the spectral density of the current-current correlator  
at $\k=0$
\begin{equation}
    \frac{ \rho_\JJ(0, \omega)}{\omega} \equiv 
\frac{1}{\pi} \,\frac{ \mbox{Im}\,\chi_\JJ(0,\omega) }{\omega}  =
          \chi_s\,\frac{T}{M} 
\frac{1}{\pi}\, \frac{ \eta }{\omega^2 + \eta^2}\, ,
\label{spf_low}
\end{equation}
\item[2.]
The typical relaxation time of a heavy quark
is set by the inverse drag coefficient, $1/\eta = D\, (M/T)$. 
The typical distance that a heavy quark moves over the relaxation is
$\sqrt{T/M}/\eta=D\sqrt{M/T}$. The correlator 
$\chi_\NN$ is a function of a scaled spatial momentum $\bar{k}=\k D\sqrt{M/T}$ and a scaled frequency $\bar{\omega} =\omega D\,(M/T)$. 
In Fig.~{\ref{spectraljj}} we show the spectral weight
of the current-current correlator. For comparison 
we also show the free current-current correlator 
from \Eq{free_rhojj_k}.
\begin{figure}
\begin{center}
\includegraphics[height=4.5in,width=4.5in]{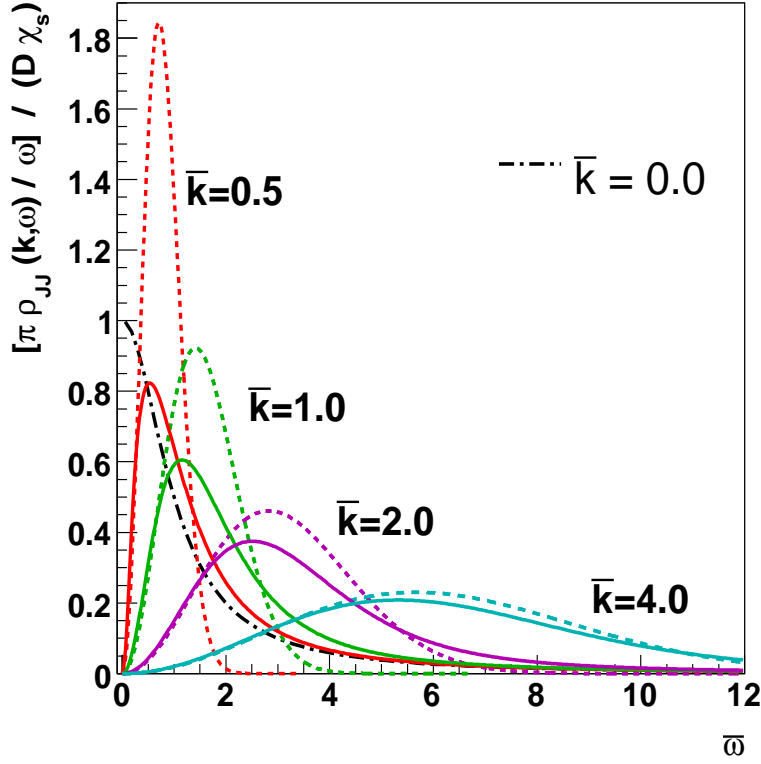}
\caption{
The spectral density of the longitudinal current-current correlator $\pi \rho_\JJ(\k,\omega)/\omega$ divided by 
$D \chi_s(\k)$  
as a function of a scaled frequency $\bar{\omega}\equiv\omega
D\,(M/T)$ for various values of a scaled momentum  
$\bar{\k} \equiv \k D\sqrt{M/T}$. The  solid lines
show the spectral density from the Langevin equations for non-zero $\bar{\k}$.
For comparison, the dotted lines show the spectral function of the 
free theory, \Eq{free_rhojj_k}, expressed in the same $\bar{\k}$ and $\bar{\omega}$ of the interacting theory.   
The dash-dotted line shows the $\k=0$ result of the
Langevin equations, \Eq{spf_low}.
}
\label{spectraljj}
\end{center}
\end{figure}

\item[3.] Noting that $\chi_\NN(\k,\omega) =-\int_{0}^{\infty} e^{i\omega t} \chi_s \partial_t P(\k,t)$ with 
$P(\k,t)=e^{-k^2 \sigma^{2}(t)/2}$, it is easy to verify the 
consistency relation $\chi(\k, 0) = \chi_s(\k)$.

\end{itemize}

\section{Numerical Estimate of the Euclidean Correlator}
\label{numerics}

In this section we will give a numerical estimate of the 
Euclidean vector current correlator.  We will
parametrize the spectral density with low 
and high frequency  contributions. 
\st
   \rho_\JJ(\k, \omega) = \rho_\JJ^{\rm low}(\k,\omega) + \rho_\JJ^{\rm high} (\k,\omega)\,.
\stp

The high frequency part is present at zero temperature 
and will be   parametrized as a  $J/\psi$ 
resonance plus a continuum
\begin{equation}
\rho_\JJ^{\rm high}(\k=0,\omega)=\Mj^2 \Fv^2 \,\delta(\omega^2-\Mj^2)
+\frac{N_c}{8 \pi^2}\,\theta(\omega^2-4 M_{D}^2)\, \omega^2 \sqrt{1-\frac{4 M_D^2}{\omega^2}}\,
\left(\frac{2}{3}+\frac{4 M_D^2}{3\omega^2}\right) \, .
\label{spf_ansatz}
\end{equation}
Here $\Fv$ is the $J/\psi$ coupling to dileptons as described
in \app{resonance_rho}. The continuum contribution is motivated by 
the free spectral function calculated in \app{loopspec}, but
we have replaced
$2 M$ with the open charm threshold $2 M_D$. 

For the low frequency part of the spectral 
function we will take two functional forms. The
first form is the Lorentzian from the Langevin equations 
\st
          \frac{\rho_\JJ(\k=0,\omega)}{\omega} = \chi_s\,\frac{T}{M} 
\frac{1}{\pi}\, \frac{ \eta }{\omega^2 + \eta^2}\, ,
\label{lorentzian_ansatz}
\stp
where $\eta=\frac{T}{MD}$.
This form is rigorously true when
$\frac{T}{MD} \ll T$, and the frequency small $\omega \lsim \eta \ll T$.

These inequalities are strained in our numerical 
work. For instance, for $T/M_c \approx 1/5$  and $D \sim 0.25/T$,
$\frac{T}{MD}$, is not really much less than $T$.
Further, as discussed in \Sect{transport_in_euclid},
the transport contribution is dominated
by the second moment of the spectral function
\st
    \int \frac{d\omega}{\omega} \,\rho_{JJ}(\omega)\,\omega^2 \, .
\stp
For the Lorentzian, this moment diverges and  the
transport contribution to the correlator
is sensitive  to the high frequency behavior of the ansatz where 
the Langevin approach is not valid.  The higher 
moments open up the white noise in the Langevin equations.
We therefore
considered a Gaussian ansatz
which falls much more rapidly at infinity
\st
   \frac{\rho_{\JJ}(\omega)}{\omega} = \chi_s \frac{T}{M}\, 
\frac{1}{\sqrt{2\pi \eta^2_{\scriptscriptstyle G}} } 
e^{-\frac{\omega^2}{2\,\eta^2_{\scriptscriptstyle G}} } \,.
\label{gaussian_ansatz}
\stp
The parameter,  $\eta_{\scriptscriptstyle G} = \sqrt{\frac{\pi}{2}}\frac{T}{MD}$, is fixed from the 
relation between the spectral density and the diffusion coefficient
coefficient,
$\left. \frac{\rho(\omega)}{\omega}\right|_{\omega=0} = \frac{\chi_s D}{\pi}$.
The integral under this smeared delta function is 
again $\chi_s \, T/M$.
By comparing these functional forms we obtain a feeling
for the uncertainties our the estimate.

The temperature dependence of the Euclidean correlators comes
from two sources: the temperature dependence of the 
spectral function $\rho(\k,\omega,T)$, and the trivial
temperature dependence of the integration kernel, \Eq{spectral_rep}.
We obviously want to separate the interesting temperature dependence 
coming from the spectral function from the trivial temperature dependence coming from the integration kernel. This can be done by defining the reconstructed correlator \cite{datta04}.
\st
G^{{\rm rec}}_{JJ}(\k,\tau,T)=\int_0^{\infty}d\omega\, \rho_{JJ}(\k,\omega,T=0) \,
\frac{\cosh(\omega(\tau-1/(2T)))}{\sinh(\omega/(2T))}\,.
\label{grec}
\stp
If  the spectral function does not change above the deconfinement temperature
$T_c$, the ratio $G_{JJ}(\k,\tau,T)/G_{JJ}^{{\rm rec}}(\k,\tau,T)$ should be unity.

\begin{figure}
\includegraphics[width=10cm]{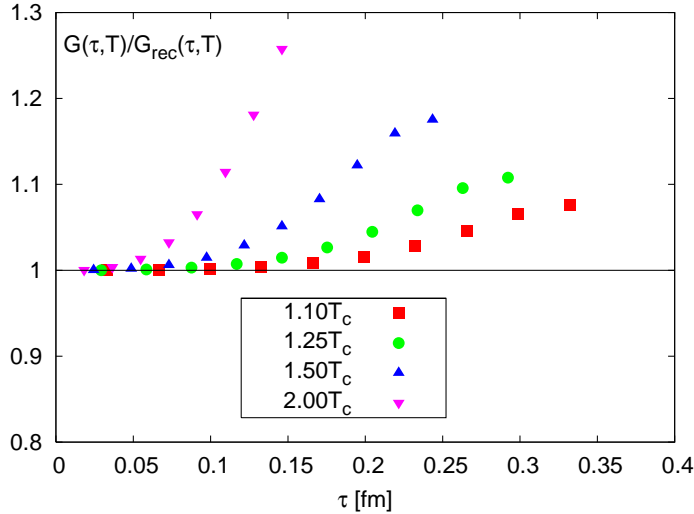}
\caption{The ratio $G_{JJ}(\tau,T)/G_{JJ}^{\rm rec}(\tau,T)$ for different
temperatures and $\k=0$.}
\label{grec_fig}
\end{figure}

First we estimate the relative importance of the transport contribution to the correlator. 
For closer comparison with existing lattice data,
we consider the diffusion
of heavy quarks in a gluonic plasma where the transition temperature is $T_c=270\,\mbox{MeV}$ 
\cite{edwinowe}.
At this stage we can set the $\eta$ to zero 
($D=\infty$) and consider only the free spectral function.
The charm quark mass $M_c$ is taken to be $1.3$ GeV.
In accord with lattice data \cite{umeda02,asakawa04,datta04}, 
we will assume that $J/\psi$ 
is not modified by the medium and 
determine $\Fv$ from its dilepton width (see \app{resonance_rho}). 
$\Mj$ and $M_D$ are taken from the Particle Data Book \cite{pdg}  
In Fig. \ref{grec_fig} we show $G_{JJ}(\k,\tau,T)/G_{JJ}^{{\rm rec}}(\k,\tau,T)$
for several temperatures. The transport contribution is 
of order $7-12\%$ and is the  
the only source of the temperature dependence seen in \Fig{grec_fig}. 
A similar enhancement was found in actual lattice calculations \cite{dattasewm04}.

Analytic understanding can be gained by 
performing the integral over the kernel at $\tau=1/(2T)$. In the
heavy quark limit, we set $\Mj\approx2M$ and $M_D\approx M$, and 
find
\begin{eqnarray*}
\label{gjjeuclid}
    \left. G_{JJ}(\k=0,\tau, T)\right|_{\tau=\beta/2}
        = 
     \underbrace{4N_c \left(\frac{M T}{2\pi} \right)^{3/2} e^{- \frac{M}{T}}\frac{T}{M}}_{\rm transport} \; + \; 
\underbrace{ M^3\left(\frac{\Fv}{2M}\right)^2 \,8\,e^{-\frac{M}{T}}}_{\rm resonance} \; + \; \\
     \underbrace{4N_c \left(\frac{M\,T}{2\pi}\right)^{3/2}  e^{- \frac{M}{T}}\left(1-\frac{T}{M}\right)}_{\rm continuum}  \,.
\end{eqnarray*}
$\Fv/2M \approx 0.131$ is small and suppresses the
resonance contribution.
The transport contribution 
is smaller by a factor of $T/M$ 
relative to the continuum contribution. 

Interactions will modify the correlator  by only a few percent.
These small changes due to the transport must 
be disentangled from other in-medium effects such as 
a small shift in the mass or width of the resonance. This can be done 
by introducing a small chemical potential for the heavy quark, $\mu_c \ll M$. Since the transport
contribution is proportional to $\chi_s$, the small
chemical potential will enhance the transport by factor of $\cosh(\mu_c/T)$, see \Eq{explicit_static}.
The small charm chemical potential will not affect the resonance and 
continuum contributions to the spectral function 
to leading order in the heavy
quark density, $\sim e^{-(M - \mu_c)/T}$.
Thus we expect
that
\begin{eqnarray}
 \delta G_{JJ} &\equiv& G_{JJ}(\tau,T,\mu)-G_{JJ}(\tau,T,0) \\
  & \simeq &
(\cosh(\mu_c/T)-1) 
\int_0^{\infty} d \omega \left. \rho_{JJ}^{\rm low}\right|_{\mu=0}(\omega)
\frac{\cosh(\omega(\tau-1/(2T)))}{\sinh(\omega/(2T))},
\end{eqnarray}
is largely insensitive to the high frequency behavior of
the spectral function.
For a thousand gauge configurations,
the statistical error in the vector current correlators can be reduced below, $0.5\%$.
One may hope that the same holds for the difference of the correlators,
$\delta G_{JJ}$. Clearly, to achieve this precision one should 
difference the two correlators before averaging over gauge configurations. This needs to studied with numerical experiments.

\begin{figure}
\includegraphics[width=10cm]{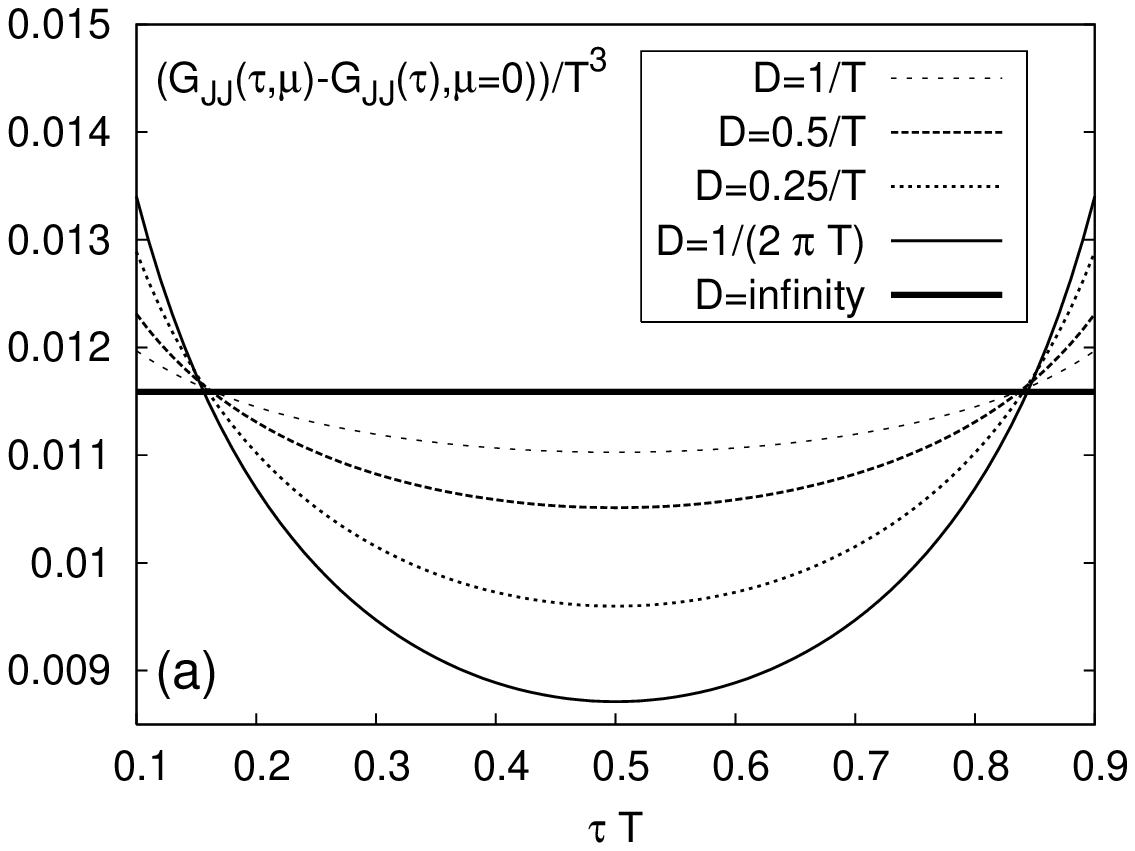}
\includegraphics[width=10cm]{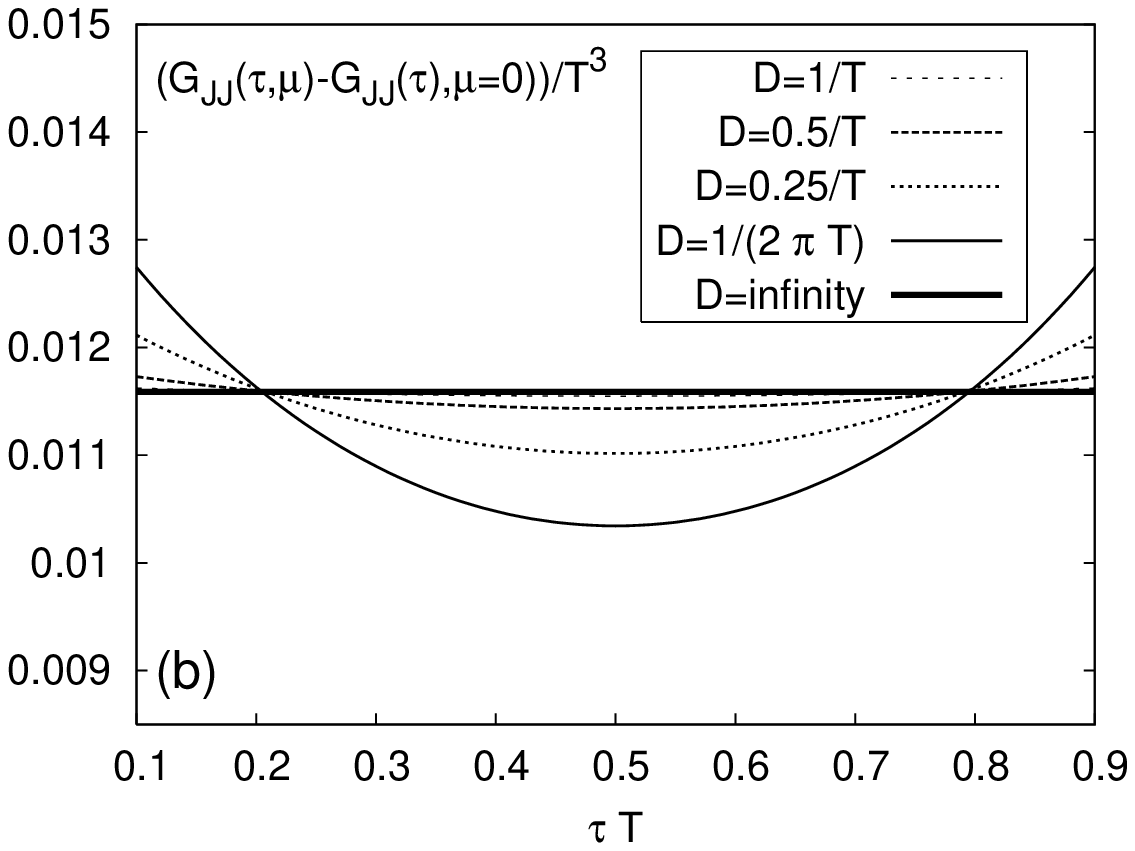}
\caption{The difference of correlators at $\mu_c=M_c/5$ and $\mu=0$ 
for the (a) Lorentzian and (b) Gaussian ans\"atze and various values
of the diffusion coefficient, D.}
\label{gmu}
\end{figure}

In Fig. \ref{gmu}(a) and (b) we show this difference for $T=1.1 T_c$, $\mu_c=M/5$  and 
different values of the diffusion constant $D$. As seen 
in Fig.~\ref{gmu}, and as expected from \Eq{GJJ},  
the effect of the diffusion coefficient 
is to provide a small curvature to the correlator and to
shift the value of the correlator downward at $\tau=\beta/2$.

First we will concentrate on the curvature.
If the final precision is 0.5\% and $D \lsim 1/(\pi T)$, 
then from Fig. \ref{gmu}(a), one could hope 
that  the curvature is
large enough to be determined in lattice simulations. 
 In practice, it will be difficult to guarantee 
that the continuum contribution will not affect the 
extracted value. 

\begin{figure}
\includegraphics[height=3.0in,width=3.0in]{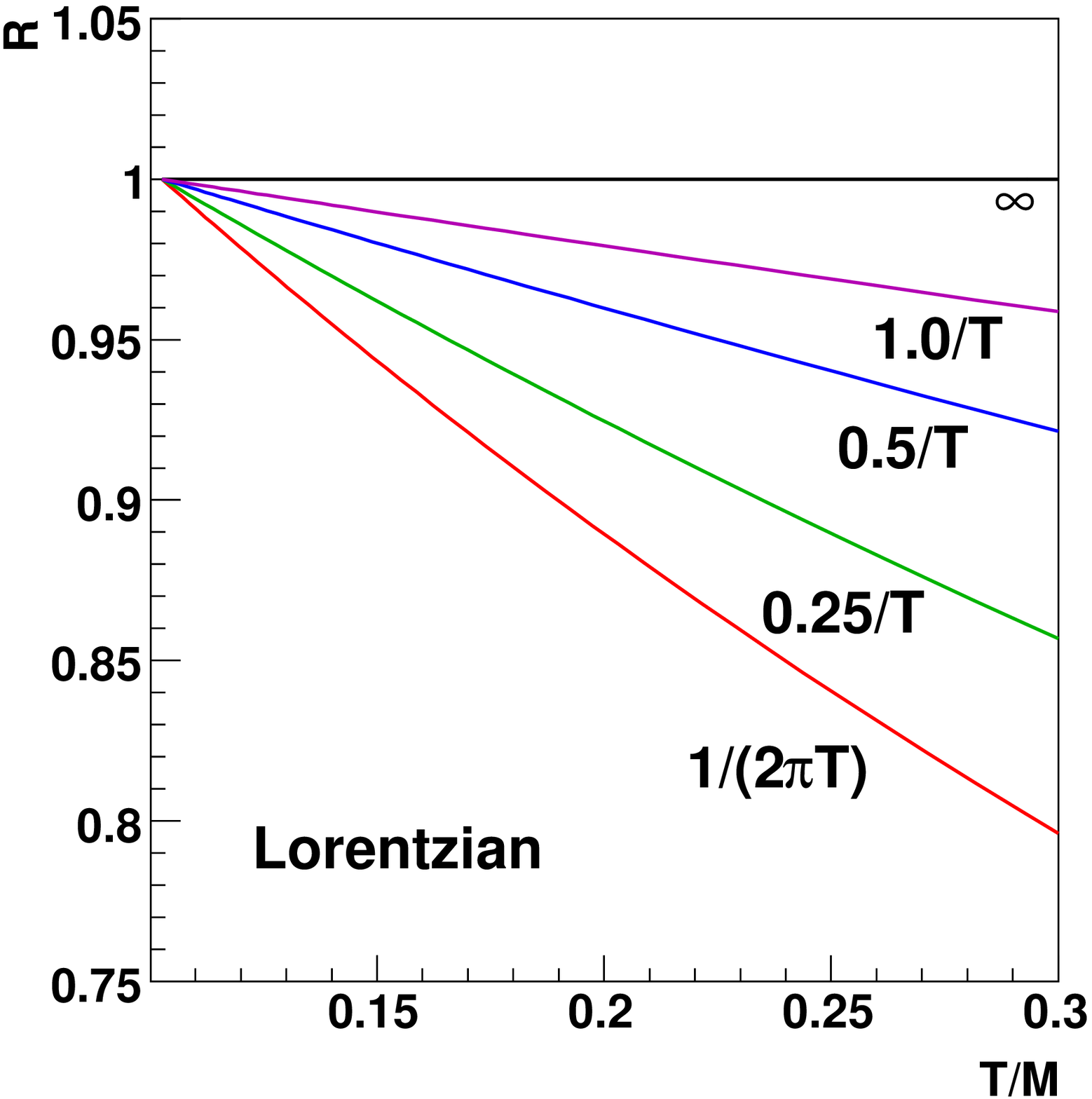}
\includegraphics[height=3.0in,width=3.0in]{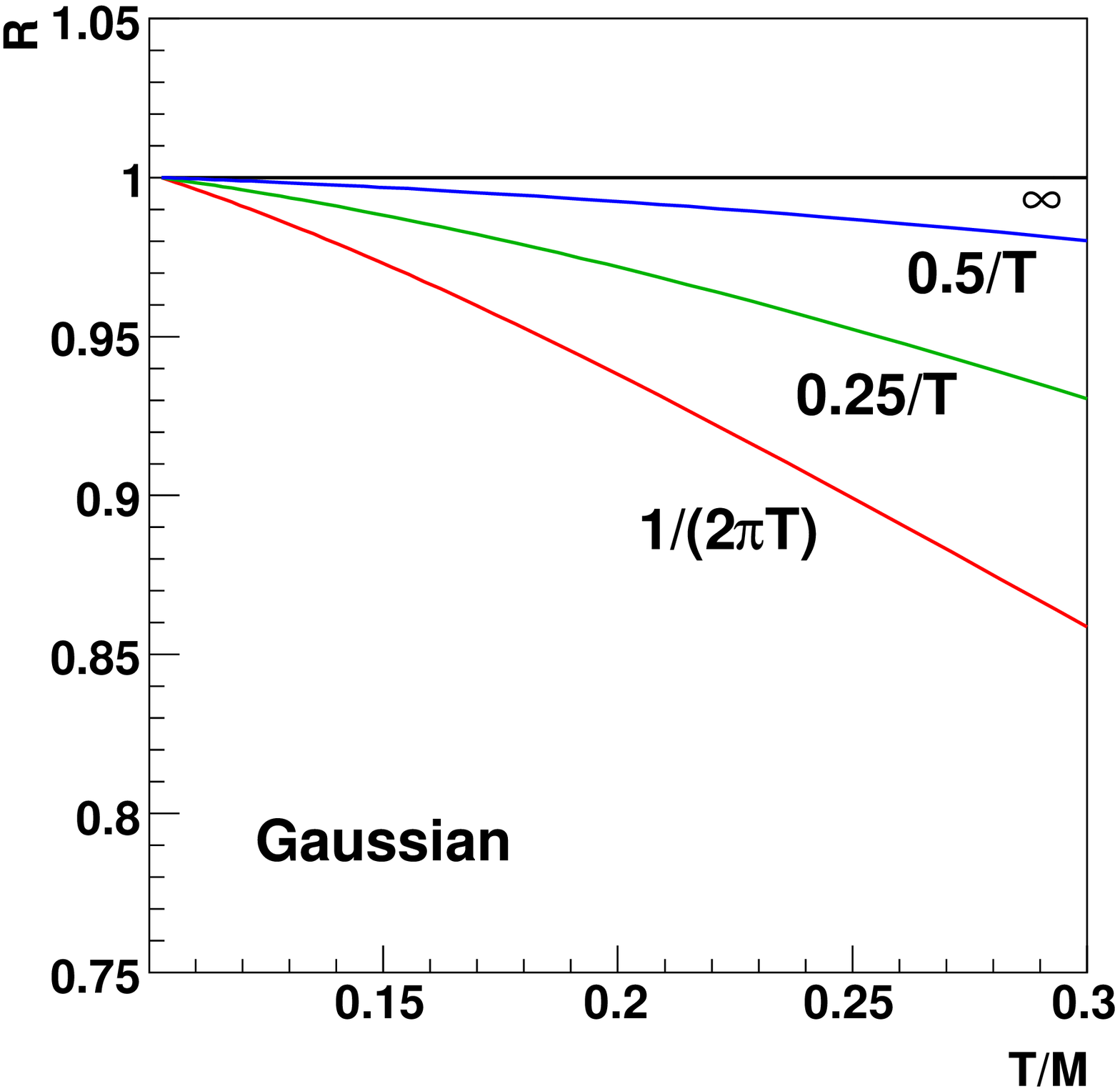}
\caption{The relative transport contribution to the correlator as
a function of quark mass, \Eq{massratio}, for  the (a)  Lorentzian and  
(b) Gaussian ans\"atze to the low frequency spectral
function. The numbers indicate the diffusion coefficient 
for each ansatz.  }
\label{massdep}
\end{figure}

The downward shift of 
the correlator at $\beta/2$ from the constant value, $\chi_s T/M$, is a much larger effect.
To isolate this transport contribution we consider the
difference, $\delta G_{JJ}(M)$, as a function of the heavy quark 
mass. We plot the ratio
\st
\label{massratio}
     R(M) \equiv \left. \frac{ \delta G_{JJ}(M)/(\chi_s(M) T/M)  }{\delta G_{JJ}(M_0) /(\chi_s(M_0) T/M_0) } \right|_{\tau=\beta/2}\,.
\stp
For the free theory this quantity is one and is independent of the
heavy quark mass.  Deviations from one are a signature of
interactions.  \Fig{massdep}(a)  and (b) show this ratio as a function 
of the heavy quark mass for the Lorentzian and Gaussian ans\"atze.
Examining \Fig{massdep}, we conclude that if the diffusion
coefficient is sufficiently small, $D\lsim 1/(\pi T)$, the transport peak
should be visible in the mass dependence of the Euclidean correlator.
Additional critical remarks are left to the conclusions.

\section{Brief Summary and Discussion}

The Euclidean current-current correlator is remarkably insensitive to 
the heavy quark diffusion coefficient. Indeed,
to leading order in $T/M$, the Euclidean current-current correlator
is independent of the diffusion 
coefficient.\footnote{This is true whenever there is a separation
between the transport and temperature time scales. Previously, Aarts
and Resco found  that Euclidean stress tensor correlations are
independent of the coupling constant to leading order \cite{aarts02}.
}
This is explained as follows (see \Sect{transport_in_euclid}). The $\tau$ derivatives 
of the euclidean current-current correlator at $\tau=\beta/2$,
the moments of the spectral function, and time derivatives 
the real-time retarded correlator at $t=0$, are in one to one correspondence.  Thus, the
value of the current-current correlator (i.e. the zero-th derivative)
is determined only by short times and may 
be calculated with the free streaming Boltzmann equation. 
In the end, the value of the current-current correlator at $\beta/2$ is
simply $\chi_s T/M$, where $\chi_s$ is the static 
susceptibility and $T/M$ reflects average thermal velocity squared.
Higher $\tau$ derivatives (or moments of the spectral density) 
reflect the width of the transport peak and contain 
useful information about the transport time scales.

In a free theory, the spectral density is proportional to 
a delta function 
\[
    \frac{\rho_{JJ}(\k=0,\omega)}{\omega} = \chi_s \frac{T}{M} \delta(\omega) \,,
\] 
which reflects the fact that in the free case, the 
spatial current is conserved  in addition to the charge.
This result may be found either by using the free 
streaming Boltzmann equation (see \Sect{transport_in_euclid}) or performing
a one loop expansion (see \app{loopspec}). 
Since the
spectral density is proportional to a delta function,
higher $\tau$ derivatives, or moments of the
spectral function, vanish and  the  Euclidean
current-current correlator is a constant, independent of $\tau$ (see also Ref.~\cite{aarts02}). 
In the 
interacting theory the delta function is smeared. 
Using the Langevin equations of motion,  
we analyze in \Sect{langevin_section}  how this delta function is smeared as a 
function of $\k$ and $\omega$.  This result together 
with the free theory is summarized in \Fig{spectraljj}.
At $\k=0$, the Langevin effective theory  dictates the replacement
\[
    \delta(\omega) \rightarrow  \frac{1}{\pi} \frac{\eta}{\omega^2 +\eta^2} \, ,
\]
where $\eta = T/(MD)$ and $D$ is the diffusion coefficient 
of the heavy quark.

With this Lorentzian form  for the spectral function at small omega, we adopted
a simple transport + resonance + continuum model for the full spectral
function and studied how the Euclidean correlator  is modified by the
transport peak in \Sect{numerics}.  We also smeared the 
delta function with a Gaussian to illuminate the sensitivity 
to the Lorentzian ansatz which is only valid in a heavy quark limit 
and for $\omega \lsim T/(MD)$. 

Generally, the transport contribution to the full correlator
is suppressed by a
factor of $T/M$ relative to the continuum contribution (see 
\Eq{gjjeuclid}). To disentangle
the transport from the continuum and resonance contributions we proposed
differencing  two current-current correlators -- one at finite heavy quark chemical
potential and  one at zero chemical potential, $\delta G_{JJ}(\tau) \equiv G_{JJ}(\tau,\mu) - G_{JJ}(\tau,0)$ . This difference is
proportional to the low frequency contribution and is independent of the high frequency contribution to leading order
the heavy quark density, $\sim e^{-(M-\mu)/T}$. 
With this procedure, 
the transport contribution can be separated from the other
contributions  at least parametrically.
In practice (as opposed parametrics) our numerical work in
\Sect{numerics} shows that extracting this piece is difficult though
not impossible.  A major unknown is 
the final precision when the difference of 
correlators  is calculated. Clearly, one should difference 
and then average over gauge configurations. Exploratory lattice studies are needed to estimate this precision.

The transport contribution to the correlator is  displayed separately
in \Fig{gmu}(a) and (b) as a function of the diffusion
coefficient.  As analyzed in \Sect{transport_in_euclid},
the effect of the diffusion coefficient is to shift the value of the
current-current correlator down from its free value $\chi_s T/M$,
and to curve the correlator at $\beta/2$.   Parametrically, these
effects are suppressed by $(T/MD)^2$ relative to $\chi_s T/M$. 
The figure illustrates that if
the diffusion coefficient is much greater than $1/T$ it will be
difficult to measure the  second derivative at $\beta/2$.  However if
the precision is $0.5\%$ it may be possible, although it will be hard to
guarantee that the continuum contribution has been completely
subtracted.
To eliminate  the continuum contribution it is desirable to make the
mass as large as possible. On the other hand, the transport signal is
proportional to $(T/MD)^2$ and therefore is suppressed by the mass. 
Ultimately, numerical experiments will determine the
optimal heavy quark mass.

Even with these complications, \Fig{gmu} shows that
the Euclidean correlator at $\beta/2$ is clearly 
shifted downward from its free value, $\chi_s\,T/M$.  This shift
also is indicative of the width of the transport peak. To evaluate
the magnitude of this shift, we proposed measuring 
\[
      \left. \delta G_{JJ}(M)/(\chi_s(M) \, T/M) \right|_{\tau=\beta/2} \,,
\]
as a function of quark mass; this quantity
is independent of the mass in the free theory. As is shown
in \Fig{massdep}(a) and (b), in 
the interacting theory the width of the transport peak 
makes this quantity mass dependent.  Judging 
from \Fig{massdep}, if the diffusion coefficient is
less than $\lsim 0.25/T$ the effects of the transport
peak should be visible in this mass 
dependence. 

Measuring \Fig{gmu} and \Fig{massdep} on 
the lattice is very difficult. The importance of 
such measurements should spur effort.
Only measurements of this kind  
can seriously challenge the strong coupling assumptions
that underly the hydrodynamic interpretation of the RHIC
results.

\noindent {\bf Acknowledgments.} We thank Guy D. Moore  
for useful discussions.  D.~Teaney was supported by grants from
the U.S. Department of Energy, DE-FG02-88ER40388 and DE-FG03-97ER4014.  
P.~Petreczky was also supported by the U.S. Department of Energy,  DE-AC02-98CH1086. He is a Goldhaber and RIKEN-BNL fellow.

\appendix
\section{Diffusion of a Brownian Particle}
\label{brownian}

The goal of this appendix is to determine the 
probability $P(\x,t)$ that a
Brownian particle will move a distance $\x$ from the origin over
a time $t$. Consider the 
discretized Langevin equations: 
\begin{eqnarray}
    \x_{t+1} - \x_t  &=& \frac{\p_t}{M}\,, \\
    \p_{t+1} - \p_t  &=& -\eta \p_t\, \Delta t + {\bf \xi}_t \,,\qquad 
  \llangle \xi_t^i \xi_{t'}^{j} \rrangle = \frac{\kappa}{\Delta t} \delta^{ij} \delta_{tt'}  \,,
\end{eqnarray}
where the noise is drawn from a Gaussian distribution with 
the specified variance.

Let $W[\p_0,\p_1, \dots, \p_n]$ be the probability of
having a sequence of momenta, $\p_0, \p_1, \dots, \p_N$,
where $\p_0$ is the momentum at time zero and $\p_N$ is the 
momentum after $N$ time steps.
The probability of having momentum $\p_0$ is 
given by the thermal distribution
\[
     P(\p_0) = \frac{e^{-\frac{p^2_0}{2\,M\,T}}}{(2\pi MT)^{d/2}} \, .
\]
Here and below $d=3$ is the number of space dimensions.
The probability to have momentum $\p_1$ given $\p_0$  
is the probability that the noise will attain the appropriate
value 
\[
   P(\p_1 | \p_0) = \int d^d\xi \,\delta^d(\p_1 - (\p_0 - \eta\,\p_0\Delta t + {\bf\xi}\Delta t) )
\left(\frac{\Delta t}{2\pi \kappa} \right)^{d/2} e^{-\frac{\Delta t}{2\kappa} \xi^2 } \, .
\]
Continuing in this way we deduce that probability distribution 
is
\st
\label{Wprob}
W[\p_0,\p_1,\dots,\p_n] = 
\frac{
   e^{-\frac{p_0^2}{2 M T} }
     }{(2\pi M T)^{d/2}}
 \frac{1}{(2\pi\kappa \Delta t)^{N\,d/2}} \, 
 \exp\left(-\sum_{i=0}^{N-1} 
 \frac{\Delta t}{2\kappa} (\dot{\p}_i + \eta \p_i)^2  \right) \, .
\stp
where $\dot{\p_i} = (\p_{i+1} - \p_i)/\Delta t$.

Now the probability to move a distance $\Delta x$ 
over a time $\Delta t$ can be written as 
\st
\label{Peqn}
 P(\Delta x, \Delta t) = \int \prod_{i=0}^{N} d^d\p_i \,W[\p_0,\p_1,\dots,\p_N] \,
\delta^{d}(\Delta x - \sum_{i=0}^{N-1} \frac{\p_i}{M} \Delta t) \, .
\stp 
We now rewrite the delta function as a Fourier integral 
and substitute \Eq{Wprob}  into 
\Eq{Peqn} to obtain
\begin{eqnarray}
\label{Peqn2}
   P(\Delta x, t) &=& \int \frac{d^d\k}{(2\pi)^d}  \,
   \prod_i^{n} d^d\p_i \, e^{i \k \cdot \Delta \x} \,
   \frac{ 
      e^{-\frac{p_0^2}{2 M T} }
       }
     {(2\pi M T)^{d/2}}
 \frac{1}{(2\pi\kappa \Delta t)^{Nd/2}} \,  \times \nonumber \\
 & & \times  \exp\left(-i\sum_{i=0}^{N-1} \frac{\Delta t}{M}\k\cdot\p_i - \sum_{i=0}^{N-1}
 \frac{\Delta t}{2\kappa} (\dot{\p}_i + \eta \p_i)^2 
\right) \,.
\end{eqnarray}
The integrals in \Eq{Peqn2} are all Gaussian and can be performed.
We performed the integrals in reverse order, $\p_n,\p_{n-1},\dots,\p_1, \p_0$ and
finally the $\k$ integral. 
The result is a Gaussian
\[
   P(\Delta x, t) = \frac{1}{(2\pi \sigma^2)^{d/2}}\, e^{-\frac{1}{2} \, \frac{(\Delta x)^2}{\sigma^2} } \,, 
\]
with width
\[
  \sigma^2 = \frac{T}{M} I_1^2 +  \frac{\kappa}{M^2}I_2 \,, 
\]
where  the discretized integrals $I_1$ and $I_2$ are,
\begin{eqnarray*}
   I_1 &=& \Delta t \sum_{i=0}^{N-1} (1-\eta \Delta t)^i 
        \longrightarrow \int_0^{t} dt' e^{-\eta(t -t')}\,, \\
   I_2 &=& (\Delta t)^3 \,  
 \sum_{i=0}^{N-1} \left[ \sum_{j=0}^i (1-\eta \Delta t)^j\right]^2
        \longrightarrow \int_0^{t} dt' \left[ \int_{t'}^{t} dt^{''} 
e^{-\eta(t-t^{''}) } \right]^2 \,.
\end{eqnarray*}
Performing the continuum integrals, and liberally using the 
relations $D=\frac{T}{M\eta}=\frac{2 T^2}{\kappa}$,
yields our final continuum form for the 
width:
\st
\label{width}
  \sigma^2(t) = 2 D t - \frac{2D}{\eta}(1 - e^{-\eta t} ) \,.
\stp
For large times, we have  $\sigma^2(t) \approx 2 D \,t$ as expected
from the ordinary diffusion equation. For small times, we 
have $\sigma^2(t)\approx (T/M)\, t^2$ reflecting the initial 
thermal distribution of heavy quarks, $\llangle v^2/3\rrangle=T/M$.

\begin{section}{The Free Spectral Function}
\label{loopspec}

To evaluate the high frequency behavior of the spectral 
function  let us evaluate the free spectral function using standard methods \cite{leBellac}.
To this end we will calculate Matsubara correlator
\st
 G_E^{\mu\nu}(\k,k_4) = \int_0^{\beta} d\tau \int d^3\x \>e^{-ik_4 \tau - \k\cdot \x}\, \llangle J_{E}^{\mu}(\x,\tau) J_{E}^{\nu}(0,0) \rrangle 
\,,
\stp
with $k_4 \equiv k_E^{0} = 2\pi n T$.  With this definition of  
the Matsubara propagator
the real time retarded propagator can be determined from its Euclidean
counter part through the relation
\st
\label{continue}
 \chi^{\mu\nu}(\k,k^0) 
= (-i)^r G_{E}^{\mu\nu}
(\k, -ik_4\rightarrow  k^0 + i\eta) \,,
\stp
where $r=\delta_{\mu0} + \delta_{\nu0}$ is the number of zeroes in the indices ${\mu,\nu}$. In the notation of the 
rest of the paper $\chi_\NN(\k,\omega)=\chi^{00}(\k,\omega)$ and
$\chi_\JJ(\k,\omega) = \hat{k}^i\hat{k}^j \chi^{ij}(\k,\omega)$.

\begin{figure}
\begin{center}
\includegraphics[height=2.5in,width=4.0in]{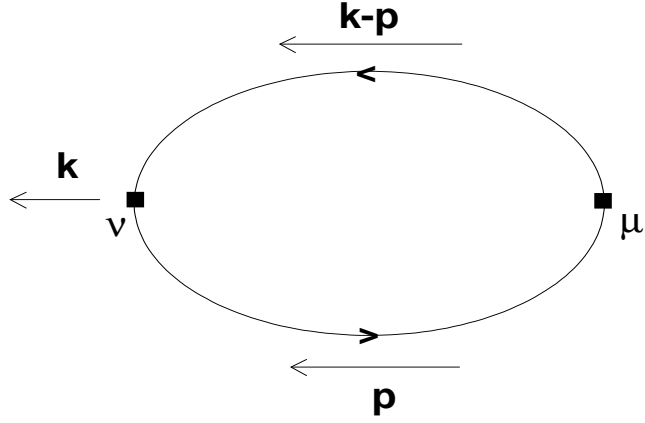}
\caption{Feynman graph contributing to the free spectral function}
\label{loop}
\end{center}
\end{figure}
The one loop contribution to the spectral function
is shown in \Fig{loop}.
\st
\label{loops}
  G_{E}^{\mu\nu}(\k,k_4) = \Nc \, T\sum_{p_4}\int \dpslash
              (-1)\, \tr\left[\frac{(-\nott{p} + M)}{p_4^2 + E_\p^2}
 \,\gamma^\mu_E\,
\frac{(\nott{k} - \nott{p} + M)}
{(k_4 - p_4)^2 + E_{\k-\p}^2 }\, \gamma^{\nu}_{E} \right] \,.
\stp
Here indices are raised and lowered with the metric tensor $g_{E}^{\mu\nu} = \mbox{diag(-1,-1,-1,-1)}$.
$\gamma_{E}^{\mu}$ satisfies $\left\{\gamma_{E}^\mu, \gamma_E^\nu\right\} = 2g_{E}^{\mu\nu}$ and 
$\nott{p} = p_{\mu} \gamma_E^{\mu} = -p^{0}\gamma^{0}_E - p^i\gamma^i_E$.

Let us examine a typical term in \Eq{loops}
\st
I_n(\k,-ik_4) = T\sum_{p_4}\, p_{4}^n\,  \frac{1}{p_4^2 + E_\p^2} \,\frac{1}{(k_4 - p_4)^2 + E_{\p -\k}^2} \; ,
\stp
where $n=0,1,2$.
Performing the frequency sum \cite{leBellac} we have,
\begin{eqnarray}
\label{sums}
I_n(\k,-ik_4)  &=&  \frac{-1 (+iE_\p)^n}{4E_\p E_{\p -\k} } \, \left[  
\frac{1 - n_\p - n_{\p -\k}} {-ik_4 - E_\p - E_{\p-\k} } - 
\frac{(-1)^n(1 - n_\p - n_{\p -\k})} {-i k_4 + E_\p + E_{\p -\k} } \right. \nonumber \\   
 & & +  \left. \frac{n_\p - n_{\p - \k}}{-ik_4 - E_\p + E_{\p - \k}}  - \frac{(-1)^n(n_\p - n_{\p - \k})}{-ik_4 + E_\p  - E_{\p - \k}} 
\right] \, .
\end{eqnarray}
Evaluating the correlator in \Eq{loops} involves performing 
the trace, evaluating the frequency sums with \Eq{sums}, and
performing the continuation $-ik_4\rightarrow k^{0} + i \eta$
as indicated by \Eq{continue}. The only 
contribution to the imaginary part of the correlator comes from 
energy denominators. In \Eq{sums} for example, the imaginary part of
a typical energy denominator after the continuation $-i k_4\rightarrow k^{0} + i \eta$ is
\[
    \mbox{Im} \frac{ -1}{(k^0+i\eta) - E_\p - E_{\p - \k}} = \pi \delta(k^0 - E_\p - E_{\p - \k} ) \, .
\]
With this identity we have
\begin{eqnarray}
\frac{\Im \chi^{00} (\k,k^0)}{\pi} &=& \int \dpslash \frac{\Nc}{4E_\p E_{\p-\k}}  
\left[ (4 E_\p \,k^0 ) D_{+} \; +\;  (-8 E_\p^2 + 4\,\p\cdot\k) D_{-} \right]\,, \\
\frac{\Im \chi^{ij} (\k,k^0)}{\pi} &=& 
\int \dpslash \frac{\Nc}{4E_\p E_{\p-\k}} 
\left[ 
(4p^ik^j + 4k^ip^j - 8p^ip^j - 4\p \cdot \k\, \delta^{ij})\,D_{+}  \right. \nonumber \\
  & &   \qquad \qquad \qquad \qquad\qquad\qquad \left. \; + \;(4E_\p k^0)\, \delta^{ij}\, D_{-} 
\right] \,,
\end{eqnarray}
where the even and odd functions $D_\pm$ are 
\begin{eqnarray}
D_{\pm} =  (1-n_\p - n_{\p-\k})\left(\delta(k^0 - E_\p - E_{\p -\k}) 
\pm \delta(k^0 + E_\p + \E_{\p-\k} \right)  \nonumber \\
           \;+ \;  (n_\p - n_{\p - \k}) \left( \delta(k^0 - E_\p + 
E_{\p-\k}) \pm  \delta(k^0 + E_\p - E_{\p - \k}) \right) \, .
\end{eqnarray}
The first pair delta functions can only be satisfied when  $|k^0|$ is
large $|k^0|\sim 2\,M$.  The second pair of delta functions
can be satisfied when $|k^0| \sim \k$. Thus
for $\k \ll T$,  the full correlator can be written as 
a sum of high and low frequency contributions
\[
   \frac{\Im \chi^{\mu\nu}(\k,k^0)}{\pi} = 
   \left[\frac{\Im \chi^{\mu\nu}(k^0\,\k)}{\pi} \right]_{\rm low}
+
   \left[\frac{\Im \chi^{\mu\nu}(k^0\,\k)}{\pi} \right]_{\rm high} \,.
\]

First let us focus on the high frequency contribution
to the spectral density. To reach an analytic expression
for the spectral density we set $\k=0$. Then the integral
over $\left(\delta(k^0 - 2E_\p) \pm \delta(k^0 + 2E_\p)\right)$ 
is easily performed, yielding
\begin{eqnarray}
 \left[\frac{\Im \chi^{L}_\JJ(\k=0,\omega)}{\pi}\right]_{\rm high} &=& 
\left[\frac{1}{3} \frac{\Im \chi^{ii}(\k=0,\omega)}{\pi}\right]_{\rm high}  
\,, \nonumber \\  
&=&\frac{\Nc\,\omega^2}{8\pi^2}
\sqrt{1 - \frac{4M^2}{\omega^2}}
\left(\frac{2}{3} + \frac{4 M^2}{3 \omega^2}\right)
\tanh\left(\frac{\omega}{2T}\right) \,,
\end{eqnarray}
This agrees with an earlier calculation \cite{Karsch:2000gi}   
after accounting for a factor of two which results from 
a sum over two flavors in that calculation.

Next we consider the low frequency contribution to 
the correlator which comes from difference of 
energies, $\delta(k^0 - E_\p + E_{\p-\k})$. 
For $\k \ll T$ we expand to first order,
\[
   n_{\p} - n_{\p -\k} \approx  
-\left(- \dndep \right) \k \cdot \v_{\p} \,,
\]
with $\v_\p=\p/E_\p$. Then the spectral density is
\begin{eqnarray}
\left[\frac{\Im \chi^{00}(\k,k^0)}{\pi}\right]_{\rm low} = 
\int \dpslash \frac{\Nc}{4E_\p^2}
\left\{ -4 p^0 k^0\, \left(-\dndep\right) \k\cdot \v_\p 
\left[\delta(k^0 - \k \cdot \v_\p) 
+ \delta(k^0 + \k\cdot \v_\p)  \right]  \right. 
\nonumber \\
+ \left. 8 E_p^2\, \left(-\dndep\right)\k\cdot \v_\p\left[\delta(k^0 - \k \cdot \v_\p) -
\delta(k^0 + \k \cdot \v_\p) \right] \right\} \,. \nonumber 
\end{eqnarray}
Integrating over $\cos(\theta_{kp})$ eliminates the
combination of delta functions symmetric with respect
$\cos(\theta_{kp})$. Integrating the anti-symmetric
combination of delta functions yields a factor of two and 
therefore
\st
\label{loop_transport}
  \left[\frac{\Im \chi^{00}(\k,\omega)}{\pi}\right]_{\rm low} = \int \dpslash 4\Nc 
\left(-\dndep\right)\, \k\cdot\v_\p \,\delta\left(\omega  - \k\cdot \v_\p\right) \,.
\stp
\Eq{loop_transport} is identical with the correlator
deduced from the free streaming Boltzmann equation.  

This expression for the retarded correlator is 
readily simplified in the non-relativistic limit 
where $n_\p = \exp\left( -p^2/(2 M T) \right)$.
The delta function  can be written as
\st
  \k\cdot\v_\p\, \delta(\omega - \k\cdot\v_p) = \frac{\omega M}{k p}\,
\delta\left(\cos\theta_{pk} - \frac{\omega M}{kp}\right) 
\Theta\left(p - \frac{\omega M}{k}\right) \,.
\stp
Integrating \Eq{loop_transport} we find 
a Gaussian with a width that is proportional to $k^2$,
\st
 \left[\frac{\Im \chi^{00}(\k,\omega)}{\pi}\right]_{\rm low} = \chi_s 
    \,\omega\ \frac{
1} {\sqrt{2\pi k^2 \llangle \frac{v^2}{3} \rrangle } }
\exp\left(-
\frac{\omega^2}{2k^2\llangle \frac{v^2}{3}\rrangle}\right)  
 \,. 
\stp
Here, $\llangle v^2/3 \rrangle=T/M$ and $\chi_s$ is the static susceptibility in the non-relativistic
limit, \Eq{explicit_static}.
In the limit that $k\rightarrow 0$ the width of the Gaussian
approaches zero and  we have
\st
 \left[\frac{\Im \chi^{00}(\k=0,\omega)}{\pi}\right]_{\rm low} = \chi_s
    \,\omega \delta(\omega) \,.
\stp
With this knowledge and the relation between the density-density and current-current correlators \Eq{chiJJ},  we find $\chi_\JJ$
\st
\label{free_rhojj_k_app}
 \left[\frac{\Im \chi_\JJ^L(\k,\omega)}{\pi}\right]_{\rm low} 
=  \chi_s \, 
    \frac{\omega^3}{k^2}\frac{
1} {\sqrt{2\pi k^2 \llangle \frac{v^2}{3} \rrangle } }
\exp\left(-
\frac{\omega^2}{2k^2\llangle \frac{v^2}{3}\rrangle}\right)   \,,
\stp
In the limit that $k\rightarrow0$ this function also approaches
$\omega \delta(\omega)$
\st
\label{free_rhojj_k0_app}
 \left[\frac{\Im \chi_\JJ^L(\k,\omega)}{\pi}\right]_{\rm low} 
   =  \chi_s \llangle \frac{v^2}{3} \rrangle  \,\omega \delta(\omega) \,.
\stp

\end{section}

\begin{section} {Resonance Spectral Function}
\label{resonance_rho}
The coupling of a $J/\psi$ to the electromagnetic
current at $T=0$ can be written as
\st
\label{Fdef}
   \llangle 0 | J^{\mu}_{EM}(0) |\p,\sigma \rrangle 
     = eQ \, \Fv \Mj \, \epsilon^{\mu}_{\sigma}(\p) \, .
\stp
Here $\Mj$ is the $\jpsi$ mass, 
$J_{\EM}^\mu=e Q\,\bar{c} \gamma^\mu c\,$, $e$ the charge of the positron,
$Q=+2/3$ and $\Fv$ is the electromagnetic decay constant. 
In writing this equation we have used the fact that
$p_{\mu} \llangle 0 | J^{\mu}_{EM}(0) |\p,\sigma\rrangle$ 
vanishes by current conservation.
The decay rate of unpolarized $\jpsi$  into $e^+e^-$  may
be expressed in terms of $\Fv$: 
\st
   \Gamma(\jpsi\rightarrow e^+e^-) = \frac{4\pi}{3} 
\frac{Q^2\alpha_{\EM}^2}{\Mj} \Fv^2  \, .
\stp 
Using the Particle Data Book \cite{pdg} we obtain, $\Fv/\Mj=0.131$. 

Using \Eq{spectral_density}, \Eq{Dgreater}, and \Eq{chiL}, the spectral density at $\k=0$ can be written as follows:
\st
\rho^L_\JJ(\k=0,\omega) = \frac{1}{2\pi}\left[ \frac{ D^>_{ii}(\k,\omega) }{3} - 
\frac{D^{<}_{ii}(\k,\omega)}{3} \right] \, ,
\stp
where $D^>_{ii}(\k,\omega)$ is 
\st
 D^>_{ii}(\k,\omega) = \int d^4x\, e^{+i\omega t - i\k\cdot\x} 
\llangle J^{i}(x)\, J^i(0) \rrangle  \, .
\stp 
Here the averages denote thermal averages  and 
$J^\mu(x) \equiv \bar{c} \gamma^\mu  c$.  We will
assume that the $\jpsi$ coupling and mass are independent 
of temperature and simply replace the
thermal average with vacuum averages.
In the frequency domain of the resonance we may assume
that one particle intermediate $\jpsi$ states dominate the correlator. 
Inserting one particle states we find 
\st
  D^>_{ii}(\k,\omega) = \sum_{\sigma} \int \frac{d^3\p}{2E_\p (2\pi)^3} \int d^4x\, e^{+i\omega t -i\k\dot\x}
\llangle 0 | J^{i}(x) |\p\sigma\rrangle 
\llangle \p\sigma | J^i(0)| 0\rrangle  \,.
\stp
Using translation invariance,  
$
\llangle 0 | J^{i}(x) | \p\sigma\rrangle = e^{-ip\cdot x} 
\llangle 0 | J^{i}(0) | \p\sigma\rrangle 
$, we perform the momentum and space-time integrals  and
find
\st
  D^>_{ii}(\k,\omega) = \frac{2\pi}{2E_k} \delta(\omega - E_k) \,
  \sum_{\sigma} \llangle 0 | J^i(0) |\k\sigma \rrangle \llangle 
\k\sigma | J^i(0) |0\rrangle \,.
\stp
We now specialize to $\k=0$ and use \Eq{Fdef} to obtain 
\st
  \frac{ D^>_{ii}(0,\omega)}{3} = \frac{2\pi}{2\Mj} \delta(\omega - \Mj) \, \Fv^2 \Mj^2 \,.
\stp
A similar calculation yields $D^<_{ii}(0,\omega)$  and the 
resonance contribution to the spectral function reads
\st
  \rho_\JJ(0,\omega) = \frac{ \Fv^2 \Mj^2}{2\Mj} \left[ \delta(\omega - \Mj)  - \delta(\omega + \Mj) \right] \, .
\stp
 
\end{section}

\end{document}